# Origins of thermal spin depolarization in half-metallic ferromagnet CrO$_2$

Hirokazu Fujiwara[1*], Kensei Terashima[2], Masanori Sunagawa[1], Yuko Yano[1], Takanobu Nagayama[1], Tetsushi Fukura[1], Fumiya Yoshii[1], Yuka Matsuura[1], Makoto Ogata[1], Takanori Wakita[2], Koichiro Yaji[3], Ayumi Harasawa[3], Kenta Kuroda[3], Shik Shin[3], Koji Horiba[4], Hiroshi Kumigashira[4,5], Yuji Muraoka[1,2], and Takayoshi Yokoya[1,2]

[1]*Graduate School of Natural Science and Technology, Okayama University, Okayama 700-8530, Japan*
[2]*Research Institute for Interdisciplinary Science, Okayama University, Okayama 700-8530, Japan*
[3]*Institute for Solid State Physics, The University of Tokyo, Kashiwa, Chiba 277-8581, Japan*
[4]*Photon Factory, Institute of Materials Structure Science, High Energy Accelerator Research Organization (KEK),1-1 Oho, Tsukuba 305-0801, Japan*
[5]*Department of Physics, Tohoku University, Sendai, 980-8577, Japan.*

*Correspondence to: fujiwara.h@s.okayama-u.ac.jp

(Dated: December 19, 2018)

**Abstract**:

Using high-resolution spin-resolved photoemission spectroscopy, we observed a thermal spin depolarization to which all spin-polarized electrons contribute. Furthermore we observed a distinct minority spin state near the Fermi level and a corresponding depolarization that seldom contributes to demagnetization. The origin of this depolarization has been identified as the many-body effect characteristics of half-metallic ferromagnets. Our investigation opens an experimental field of itinerant ferromagnetic physics focusing on phenomena with sub-meV energy scale.

Entirely understanding magnetic phenomena in itinerant electron ferromagnets has been one of the most significant goals of condensed-matter physics [1]. In itinerant electron ferromagnets, interaction between conduction electrons and thermal spin fluctuation is of crucial importance for understanding the physical properties. As a many-body state dominating the transport properties of ferromagnetic metals, the nonquasiparticle (NQP) state, also called the spin-polaron state, was proposed in works on electron–magnon interaction [2,3]. For investigation of the behavior of the NQP state, half- metallic ferromagnets (HMFs), which have metallic electronic structures with an energy gap at the Fermi level ($E_F$) for any one electronic spin state in the ground state, are ideal substances. This is because in half-metallic ferromagnets it is predicted that the effect of electron–magnon interaction is not masked by Stoner excitations unlike the usual itinerant ferromagnets, and it makes an observable modification of the electronic structure in the close vicinity of $E_F$ [4,5]. Although many theoretical studies on NQP were reported [6–8], only a few experimental studies suggesting the existence of NQP in HMFs have been published [9,10]. Experimental electronic-structure investigation of a HMF can reveal the behavior of the NQP state, which not only advances the formulation of many-body physics but also promotes application of HMFs for next-generation spintronic devices [11].

Chromium dioxide ($CrO_2$), which has a rutile-type crystal structure, is the simplest half-metallic oxide without carrier doping. As a notable advantage, $CrO_2$ always shows almost 100% spin polarization, which is the highest value exhibited by a candidate HMF at low temperature [12–15]. This completely spin-polarized feature is suitable for exploring the many-body state. In addition, recently, $CrO_2$ has been predicted to host triple-point fermions, as well as additional Weyl points, attracting the attention of material scientists [16]. According to magnetoresistance studies, the spin polarization drops exponentially at higher temperatures, the origin of which cannot be understood from bulk magnetization alone [17–19]. As a cause of the rapid depolarization, many-body effects have also been proposed [20]. Theoretical studies based on the dynamical mean-field theory (DMFT) showed that many-body effects broadened the bandwidth of a minority spin state above $E_F$ and that the tail of the state crossed $E_F$, which permitted spin-flip scattering of the conducting majority spin electrons [6,20].

Spin-resolved photoemission spectroscopy (SRPES) is a powerful technique to directly observe spin-polarized electronic structures and determine absolute values of spin polarization based on simple analyses. In SRPES measurements using a conventional low-efficiency spin detector, the energy resolution is set to several 100 meV, which is not sufficient to observe the many-body states. However, recently, by using a high-efficiency spin detector and high-intensity low-energy light source, energy resolutions that were almost two orders higher were achieved [21–23]. This allows us to investigate the spin-polarized fine electronic structures such as the many-body states in HMFs.

In the pioneering works of SRPES on $CrO_2$ [24,25], SRPES spectra show almost 100% spin polarization near $E_F$ at room temperature. However, the energy resolutions were not good enough to discuss a depolarization near the $E_F$ characteristic of the many-body effects.

In this article, we present the temperature dependence of the electronic structure and spin polarization of half-metallic $CrO_2$(100) epitaxial films, measured using laser-based high-resolution SRPES, in order to clarify the origin of the depolarization in the magnetoresistance. The development of a minority spin state around $E_F$ with energy on the order of 10 meV above 80 K and the corresponding spin depolarization are clearly observed. The tendency of the temperature dependence of the minority spin state is consistent with the DMFT calculations and NQP theories, which constitutes spectroscopic evidence for the many-body effect in $CrO_2$.

The $CrO_2$(100) epitaxial films grown on a rutile-type $TiO_2$ (100) substrate were prepared by a closed-system chemical vapor deposition method [26]. After the synthesis, the $CrO_2$ film was removed from the quartz tube and then immediately placed under high vacuum for SRPES measurements. During the procedure, the $CrO_2$ sample was exposed to the atmosphere for approximately three minutes.

Spin-integrated and spin-resolved photoemission spectroscopy data were acquired by the laser-based spin-resolved angle-resolved photoemission spectroscopy (SARPES) apparatus at the Institute for Solid State Physics at the University of Tokyo [23]. The apparatus was equipped with highly efficient very-low-energy electron diffraction (VLEED) spin detectors, whose effective Sherman function $S_{eff}$ was 0.25, and a hemispherical analyzer (SCIENTA- OMICRON DA30L). We used a vacuum ultraviolet (VUV) laser ($hv = 6.994$ eV) with p-light-polarization as an excitation beam. During the measurement, the instrumental energy resolution was set to 20 meV and the base pressure was kept below $1 \times 10^{-8}$ Pa. The acceptance angle of the spin detector was set to 0.7°, which reliably corresponded to approximately 1% of the Brillouin zone in the XMAR plane centered at the X point. However, our ARPES spectrum does not indicate any clear angular dependence [27]; therefore, our results are considered to be angle integrated. Calibration of $E_F$ for the sample was achieved using a gold reference. We magnetized the $CrO_2$ (100) sample along the magnetic easy axis ([001] direction) by bringing the sample close to a magnet at room temperature. The approximate magnitude of the magnetic field at the sample position was 600 Oe. The schematic views of the experimental geometry and magnetization procedure are shown in Fig. S1 [27].

From Figs. 1(a) and 1(c), up to 70 K, the spin polarization decreases in the energy range above 80 meV, with the same binding-energy dependence; this we call type I depolarization. The depolarization continues to increase at a constant rate above 80 K also, while additional variations appear in the binding-energy dependence (discussed below as type II depolarization).

According to a report in Ref. [14], type I depolarization occurs over the entire spin-polarized energy range, namely, from more than $E_B \sim 1$ eV to $E_F$. This fact indicates that type I depolarization contributes predominantly to the demagnetization of the sample, which makes us speculate that the origin of type I depolarization is the same as that of demagnetization. In an earlier report, the M($T$) curve of $CrO_2$ powders followed Bloch's $T^{3/2}$ law, and it was shown that the demagnetization was attributed to spin-wave excitation [33], as in our M($T$) curve [27]. The spin-wave excitation caused a spin-mixing effect [34]. The magnitude of the spin-mixing contribution to the density of states was estimated to be $D_\downarrow(E) \approx (M_0 - M_s(T))/(M_0 + M_s(T))D_\uparrow(E)$, where $D_{\uparrow/\downarrow}(E)$ is the majority/minority spin density of states, $M_s(T)$ is the spontaneous magnetization, and $M_0 = M_s(0)$. This indicates that $D_\downarrow(E)$ shows the same energy dependence as $D_\uparrow(E)$ and that the corresponding spin polarization decreases over the entire spin-polarized energy range at elevated temperatures, which supports the present results from our SRPES measurements. Therefore, type I depolarization can be attributed mainly to the spin-wave excitation.

Nevertheless, the spin polarization measured by the laser decreases much more rapidly than the magnetization, with increasing temperature, although ideally $P$(80 meV) in Fig. 1(d) should show the same temperature dependence as the M($T$) curve. This behavior is quite similar to that of surface magnetism [35,36]. Note that $P$(80 meV) also decreases more rapidly than what was observed in the bulk-sensitive SRPES studies [13,14]. These facts indicate that the present measurements obtained using the 7-eV laser are quite surface sensitive. The probing depth of the present measurements is seemingly in contradiction to that expected from the theoretical mean free path of photoelectrons [37]. One of the reasons for this inconsistency is because the lifetime of the photoexcited electrons is extremely short, as the final states are not in the unoccupied bulk bands; however, the surface electrons can escape from the solid surface into vacuum by the short lifetime, as reported by theoretical and experimental photoemission studies [38–40]. This point is also supported by the electronic structures predicted during the band calculation of $CrO_2$ [41]. Note that our success in observing the intrinsic electronic states of $CrO_2$, despite the surface-sensitive measurements, is because of our use of a high-quality sample with very few surface contaminants.

We also found that the spin polarization evidently drops toward $E_F$ above 80 K, which we regard as another type of depolarization, named type II depolarization. With increasing temperature, the binding-energy dependence of the spin polarization seems to change slightly at 70 K. At 80 K, the spectral shape of the spin polarization clearly bends at $E_B = 10$ meV [black arrow in Fig. 1(c)] and the spin polarization drops toward $E_F$. Such a bending shape of the spin polarization is very evident at temperatures up to 120 K. Above 150 K, the bending point shifts toward the higher-binding-energy side, and the bending structure gets broader. To eliminate the Fermi cutoff, the original curves shown in Fig. 1(a) are divided by the Fermi-Dirac (FD) function at each measured temperature convoluted with the experimental resolution in Fig. 1(b). With increasing temperature, from 20 K to 120 K, while the majority spin states do not change significantly, a finite state appears in the minority spin states at and above $E_F$. This minority state causes type II depolarization.

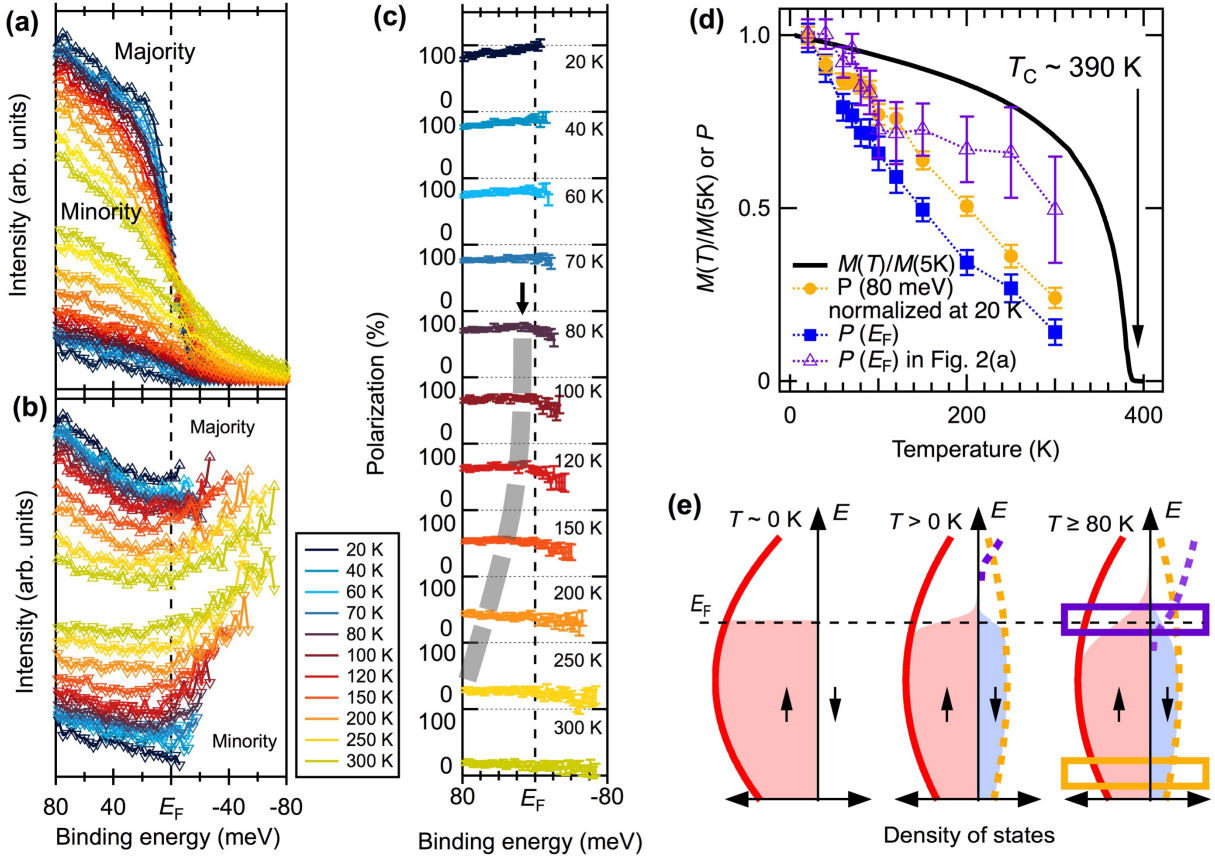

FIG. 1 Temperature-dependent spin-resolved electronic structures and corresponding spin polarizations. (a) Temperature dependence of spin-resolved spectra; (b) that divided by the FD functions adding a constant background at measured temperatures convoluted with the experimental resolution; and (c) that of the corresponding spin polarization. In (a) and (b), triangle up (down) represents the majority (minority) spin spectrum. In panel (c), the error is indicated by bars. The black arrow and gray dashed line show the bending point at 80 K and a visual guide representing the shift of the bending point, respectively. (d) Comparison of the temperature dependence of spin polarizations at various binding energies with that of the remnant magnetization along the c-axis direction for the $CrO_2$ sample, which was magnetized by a magnetic field of 1 T at 300 K. Yellow filled circles and blue filled squares show the spin polarizations at a binding energy $E_B$ = 80 meV and at $E_F$, respectively. Purple triangles show the spin polarization at $E_F$, which is normalized by its value at 80-meV binding energy at 20 K after subtracting the background from majority and minority spin spectra. The spin polarization and the spin-resolved PES spectra used to obtain the purple-triangle points are shown in Fig. 2. (e) Schematic representation of density of states at the ground state and at finite temperatures. Red and blue areas represent the occupied states. Yellow and purple dashed lines represent the minority spin states causing type I and type II depolarizations, respectively (see text). Energy regions enclosed by yellow and purple rectangles, in which the spin polarizations are averaged to obtain the temperature dependence shown in panel (d), correspond to the $P(80\ \mathrm{meV})$ and $P(E_F)$ in panel (d), respectively.

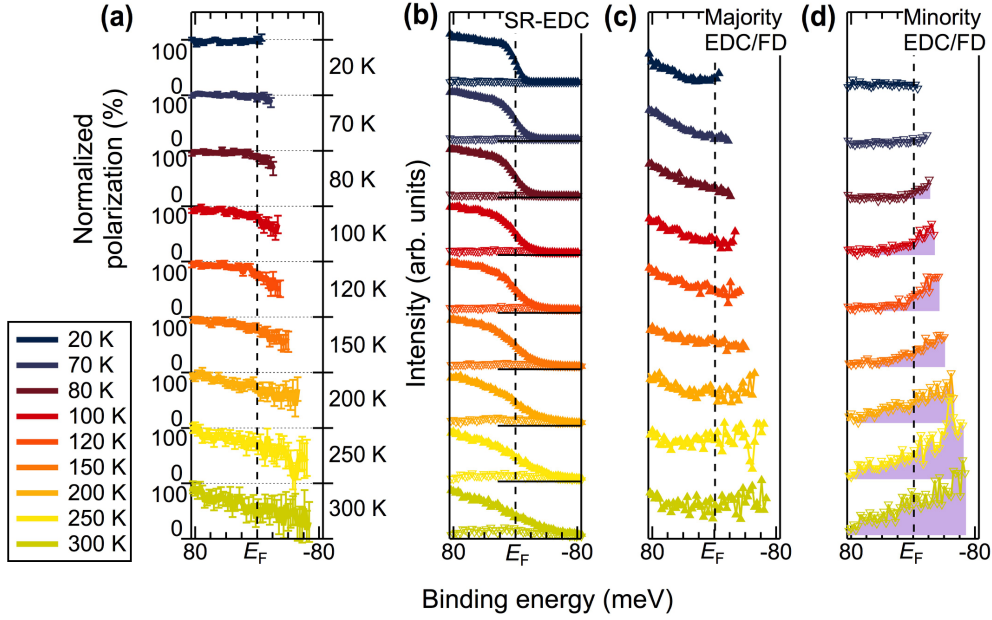

FIG. 2 Minority tail states at finite temperature. (a) Spin polarization obtained from majority and minority spin spectra subtracted background and normalized by its value at $E_B = 80$ meV for various temperatures. The error is indicated by bars. (b) Spin-resolved EDCs calculated from $I_\uparrow = (1 + P_{nor})I_{tot}/2$ (triangle-up) and $I_\downarrow = (1 - P_{nor})I_{tot}/2$ (triangle-down), where $P_{nor}$ is the normalized spin polarization shown in panel (a). (c,d) Majority and minority EDCs divided by the FD function at the measured temperature convoluted with the experimental resolution.

From Fig. 1(e), $P(80$ meV$)$ includes the contribution of only type I depolarization, as mentioned above, while the spin polarization at $E_F$, $P(E_F)$, includes the contributions of both type I and type II depolarizations. Differences between these spin polarizations appear above 80 K, and the gap becomes wider for temperatures up to 120 K. In the higher-temperature region, $P(80$ meV$)$ and $P(E_F)$ are almost parallel, and they decrease at the same rate, implying that the contribution of type I depolarization is dominant and that of type II depolarization is rather enshrouded. Actually, the majority and minority energy distribution curves (EDCs) divided by the FD function have approximately the same spectral shape above 200 K due to type I depolarization.

In order to clarify the temperature dependence of type II depolarization, we perform a background-subtraction analysis [27] and normalize the spin polarization obtained after the background-subtraction analysis by its value at $E_B = 80$ meV, as shown in Fig. 2(a). This eliminates the contribution of type I depolarization from the data of spin polarization. Similar to Fig. 1(c), the normalized spin polarization shows type II depolarization at approximately 80 K. Furthermore, type II depolarization is enhanced and broadened systematically with the increase in temperature up to 300 K. In response to the change in the spin polarization, an extended peak structure appears in the minority spin EDCs. To clarify the peak structure further, we analyze the spin-resolved EDCs divided by the FD function shown in Figs. 2(c) and 2(d). In the FD-divided minority EDCs, it is evident that a minority spin state exists at $E_F$ above 80 K (purple shaded region). Moreover, this minority tail state becomes much broader and occupies sufficiently below $E_F$. On the other hand, such considerable changes are not observed in the FD-divided majority EDCs. This fact is significant evidence that the minority tail state plays a central role in type II depolarization.

Let us discuss the origins of the minority tail state. A soft x-ray ARPES study of $CrO_2$(100) films reported that the band structure of $CrO_2$ could be understood from the local spin-density approximation with electron correlation $U$ (LSDA+$U$) calculated with effective on-site Coulomb interaction $U_{eff} \approx 1$ eV [42]. The band structure has an energy gap in the minority spin states approximately 1 eV below $E_F$ and 0.5 eV above $E_F$ [41,43,44]. Therefore, considering the energy, the minority spin bands above $E_F$ expected from LSDA+$U$ cannot be observed by broadening the FD distribution even at 300 K.

In a theoretical study [20], the minority state extending below $E_F$ was interpreted as a NQP state arising from an electron-magnon interaction. Based on NQP theories [2–6,9], the broadening of the minority spin state occurs at an energy scale on the order of the characteristic magnon energy. When the temperature exceeds the anisotropy gap in the magnon spectrum, the NQP state can exist in the occupied states. This indicates that spin polarization does not change as long as the thermal energy does not exceed the energy required to overcome the gap of the magnon spectrum; however, after it exceeds the energy, the spin polarization at $E_F$ starts to drop instantly. The characteristics of the temperature dependence are consistent with those obtained from our measurements, shown as purple triangles in Fig. 1(d). This suggests that the tail state in the minority spin state observed here can be attributed to the many-body effect. A study involving the DMFT + LSDA calculation predicted that the density of states of the minority spin above $E_F$ was broadened by the many-body effect to cross $E_F$ above 100 K [20]. This effect may cause band broadening to a larger energy extent than thermal energy. In a theoretical study, it was shown that the edge of the minority state above $E_F$ at the ground state shifted to $E_F$ and extended to $E_B \sim 100$ meV at 100 K, and then slightly shifted to the higher-energy side with more broadening at 200 K. This tendency is consistent with our SRPES results.

However, curiously, the spin polarization at $E_F$ at 200 K was higher than that at 100 K in the DMFT study [20], which does not seem to be in line with the temperature dependence of our spin polarization. In Fig. 1(d), the purple triangles fall precipitously from 70 K to 100 K, but between 100 K and 250 K, the spin polarization stays at approximately 70%. Since the trend of temperature dependence is not upward, but downward, with an increase in temperature, the temperature dependence of our spin polarization cannot be completely explained by the DMFT+LSDA calculation. Furthermore, from the DMFT+LSDA calculation, the energy scale of the occupied minority spin state is estimated to be 100 meV, which is approximately 10 times larger than that obtained by our experiment. To understand our result, further theoretical studies are required.

Here, as an alternative depolarization mechanism, we discuss the effect of a Stoner-type collapse, which may produce a spin depolarization with binding-energy dependence similar to that of type II depolarization. ARPES and SARPES studies on Ni metal reported that the exchange splitting $\Delta$ decreased from 280 meV to 130 meV with increasing temperatures from $0.47T_C$ to $0.80T_C$, and then vanished rapidly towards $T_C$ [45,46]. The temperature dependence of $\Delta$ corresponded to the magnetization curve of Ni [46,47]. In the case of $CrO_2$, the minority-spin gap above $E_F$ at the ground state can be estimated to be more than 500 meV based on ARPES and theoretical studies [41–44]. From this fact and our experimental results, the minority spin gap was reduced by as much as 500 meV with increasing temperatures up to $0.2T_C$ (~80 K), and then the gap was reduced by only 100 meV with increasing temperatures from $0.2T_C$ up to $0.75T_C$ (~300 K). This temperature dependence of $\Delta$ is different from that of Ni, which implies that type II depolarization is not attributed to the reduction of the exchange splitting.

Several studies on magnetotransport measurements sup- port our experimental results [33,48]. The study on $CrO_2$(110) films suggested that spin-flip scattering events occurred above a certain temperature $T^* = 80$ K due to electron-magnon interaction [48]. This temperature $T^*$ was in good agreement with the temperature at which type II depolarization was observed. This suggests that the electron–magnon interaction is associated with type II depolarization, supporting the NQP scenario mentioned above. Furthermore, the study in Ref. [48] reported that a crossover of the sign of the carriers from plus (holelike) to minus (electronlike) occurred at $T^*$. We observe that the minority tail state seems to move from the unoccupied side toward the higher-binding-energy side. We can speculate that the tail state is possibly an electronlike band and that it produces an electron pocket somewhere in the Brillouin zone. This variation in the electronic structure is consistent with that observed in the magnetotransport study.

We investigate the spin-depolarization process in high- quality $CrO_2$(100) epitaxial films, using high-resolution SRPES. Two types of spin depolarizations are observed. The first is type I depolarization, which develops over the entire spin-polarized energy range at a constant rate with respect to temperature. The origin of type I depolarization is attributed to spin-wave excitation. Above 80 K, type II depolarization occurs in the close vicinity of $E_F$ with an energy scale of several 10 meV. With the development of type II depolarization, a minority spin state hidden above $E_F$ at lower temperatures appeared and was enhanced in the minority spin gap with elevating temperature, while the majority spin state was not changed significantly. The temperature dependence of the minority spin state is consistent with that of the NQP theories, which constitute spectroscopic evidence for the appearance of a minority tail state attributed to a many-body effect. A thorough investigation of the fine spin-resolved electronic structure of HMFs will be important for understanding the many-body effect in itinerant electron ferromagnets and for realizing complete spin polarization at room temperature, which may accelerate the development of spintronic devices.

We acknowledge H. Adachi, S. Onari, and M. Ichioka for valuable discussions. We thank T. Matsushita and S. Toyoda for their support of our data analysis. We also thank W. Hosoda, A. Takeda, and Y. Suga for technical support, and T. Kambe, T. C. Kobayashi, and their students for their help with the magnetization measurements. The laser-based SRPES experiments were conducted at ISSP with the approval of ISSP (Proposal Nos. A296, B248, and A182). Part of the ARPES experiments were performed under the Photon Factory Proposal No. 2018G127. This work was partially supported by the Program for Promoting the Enhancement of Research Universities and a Grant-in-Aid for the Japan Society for the Promotion of Science (JSPS) Fellows (No. 16J03208) from the Ministry of Education, Culture, Sports, Science and Technology of Japan (MEXT). H. F. was supported by a Grant-in-Aid for JSPS Fellows.

# Supplemental Information for

# Origins of thermal spin-depolarization in half-metallic ferromagnet $CrO_2$

Hirokazu Fujiwara[1*], Kensei Terashima[2], Masanori Sunagawa[1], Yuko Yano[1], Takanobu Nagayama[1], Tetsushi Fukura[1], Fumiya Yoshii[1], Yuka Matsuura[1], Makoto Ogata[1], Takanori Wakita[2], Koichiro Yaji[3], Ayumi Harasawa[3], Kenta Kuroda[3], Shik Shin[3], Koji Horiba[4], Hiroshi Kumigashira[4,5], Yuji Muraoka[1,2], and Takayoshi Yokoya[1,2]

[1]*Graduate School of Natural Science and Technology, Okayama University, Okayama 700-8530, Japan*
[2]*Research Institute for Interdisciplinary Science, Okayama University, Okayama 700-8530, Japan*
[3]*Institute for Solid State Physics, The University of Tokyo, Kashiwa, Chiba 277-8581, Japan*
[4]*Photon Factory, Institute of Materials Structure Science, High Energy Accelerator Research Organization (KEK),1-1 Oho, Tsukuba 305-0801, Japan*
[5]*Department of Physics, Tohoku University, Sendai, 980-8577, Japan.*

*Correspondence to: fujiwara.h@s.okayama-u.ac.jp

# Supplementary Notes

## Supplementary Note S1. Quality of our sample surface

For carrying out the experiments, we prepared a high-quality sample on which contaminations are minimized. LEED pattern shown in Fig. S2(a) exhibits the clear rectangular-like pattern characteristics of the rutile-type (1×1) structure. The ratio of $a*/c*$ was estimated to be 0.659 which is in good agreement with that of previous studies,[13,14,28] which ensures the epitaxial growth of the $CrO_2$ film. In order to evaluate the amount of the surface contaminants on our sample, we performed valence band photoemission spectroscopy by using the He IIα resonance line ($hv$ = 40.8 eV) whose probing depth is approximately 5–10 Å.[32] The valence band spectrum shown in Fig. S2(b) has a peak located at binding energy $E_B$ = 1 eV with a clear Fermi edge. These characteristics are those of the intrinsic electronic states of $CrO_2$ according to bulk-sensitive photoemission spectroscopy studies.[13,29,37] Furthermore, we confirmed that there is the 1-eV spectral structure in spectrum measured by 7-eV laser, while the intensity of the structure is suppressed with decreasing photon energy because of decrease of photoexcitation cross-section of $3d$ bands relative to O $2p$ bands, as shown in Fig. S3. It is known that the surface of $CrO_2$ tends to change to the antiferromagnetic insulator $Cr_2O_3$, preventing from observing the intrinsic electronic structure.[30] If the sample surface is completely covered by the contaminants (mainly $Cr_2O_3$), the photoemission spectrum has a peak located at 2 eV without the Fermi edge.[30] However, no clear structure at 2 eV was observed in our measurements. Based on these facts, our sample can be regarded as a high-quality sample enough to investigate the intrinsic electronic states of $CrO_2$ by the photoemission method using the 7-eV laser.

## Supplementary Note S2. Angular dependence of photoelectrons

ARPES intensity map measured by synchrotron radiation at $hv$ = 114 eV is shown in Fig. S4. We successfully obtained clear angular dispersion in the ARPES intensity map, although we cannot observe clear angular dependence in that measured at $hv$ = 6.994 eV. Obtaining clear dispersions at $hv$ = 114 eV shows that the reason why we cannot observe any clear dispersions at 6.994 eV is not attributed to the quality of our sample surface. If our SRPES measurements by the laser are ideally "angle-resolved", we cannot observe a clear Fermi edge, because the observed area in Brillouin zone was near the zone boundary where there is no Fermi surface.[37] The reason of observation of no clear angular dependences and of Fermi edges by our SRPES measurements is probably due to the finite momentum window perpendicular to the sample surface ($k_z$ in the present study). While such a broadening of the momentum window can occur generally due to the reduction in the photoelectron escape depth as expected from the universal curve, it may not be a dominant factor for the measurement at 6.994 eV because of the longer photoelectron escape depth.[32] Therefore, we attributed this to the direct "band-gap" photoelectron emission case.[33] In this case, if an electron is excited into the band-gap region, the life time of the excited electron is much shortened, and the final state wave function consists only of surface evanescent waves which rapidly decay into the solid (Measurements become surface sensitive.). This gives rise to the widening of a momentum window perpendicular to the sample surface in an ARPES measurement, leading to obtaining information of 1 dimensional density of states rather than band dispersion. Therefore, although the acceptance angle of the spin detector was set to 0.7° in our SRPES measurements, our spin-resolved EDCs reflect the density of states of $CrO_2$.

## Supplementary Note S3. Analysis

In order to obtain absolute values of spin polarization using the VLEED detector, we used the equation $P = (1/S_{eff}) (I_+ - I_-)/(I_+ + I_-)$, where $S_{eff}$ (=0.25) is the effective Sherman function of the apparatus and $I_{+(-)}$ is the intensity of the electrons reflected by the positively (negatively) magnetized target.[31] Then, we obtain the majority ($I_\uparrow$) and minority ($I_\downarrow$) spin spectra using $I_{\uparrow(\downarrow)} = (1 +(-) P)(I_{tot}/2)$, where $I_{tot} = I_+ + I_-$. After subtracting the background from $I_\uparrow$ and $I_\downarrow$, we obtain the resulting spin polarizations from $P = (I_{\uparrow BG} - I_{\downarrow BG})/(I_{\uparrow BG} + I_{\downarrow BG})$, where $I_{\uparrow(\downarrow)BG}$ is the intensity of the majority (minority) spin electron without the background. The statistical error bars of the spin polarization were estimated from $1/(S_{eff}I_{tot}^{1/2})$.[32]

To eliminate the contribution of type I depolarization, we performed an analysis normalizing spin polarization by its value at binding energy $E_B = 80$ meV at which type II depolarization does not contribute. The normalized spin polarization is represented as $P_{nor} = (1/S_{eff}F_{nor}(T)) (I_+ - I_-)/(I_+ + I_-)$, where $F_{nor}(T)$ is the normalize factor for each temperature. The error is represented as $1/( S_{eff}F_{nor}(T)I_{tot}^{1/2})$. This error gets larger with increasing temperature, because $F_{nor}(T)$ gets smaller with increasing temperature because the unnormalized spin polarizations get smaller with increasing temperatures.

## Supplementary Note S4. Background-subtraction analysis

Since we underestimate the spin polarization when non-negligible unpolarized background exists in a SRPES spectrum, we should remove the background to obtain intrinsic values of spin polarization.[31] Blue lines with triangle-up/down in Fig. S6 show our majority and minority spin EDCs measured by the VUV laser ($hv = 6.994$ eV) at 20 K, respectively. A clear Fermi edge was observed in the majority spin spectrum while no states at $E_F$ with an energy gap of 10 meV below $E_F$ were observed. These features are just those of half-metallic ferromagnet. However, in bulk-sensitive SRPES studies, the gap size of the minority spin state is estimated to be 500 meV below $E_F$ at 40 K where the degree of spin polarization is 100 % independent on binding energy.[13,14] The present minority spin EDC and spin polarization seem to be inconsistent with those of the bulk-sensitive SRPES study. This inconsistency can be attributed to the difference of the surface sensitivity of these measurements, as mentioned in the main article. Therefore, in our measurements by the laser, we may observe non-negligible intensity from the surface contaminants in addition to that of $CrO_2$. Based on this picture, the background seen in the minority spin spectrum at 20 K in Fig. S5 can be identified to be the tail of the electronic states of the surface contaminants as $Cr_2O_3$ which has a peak at 2 eV with no Fermi edge. Therefore, we subtracted the smoothed spectrum of the minority spin one, as shown in Fig. S5, from both majority and minority spin spectra at any temperatures.

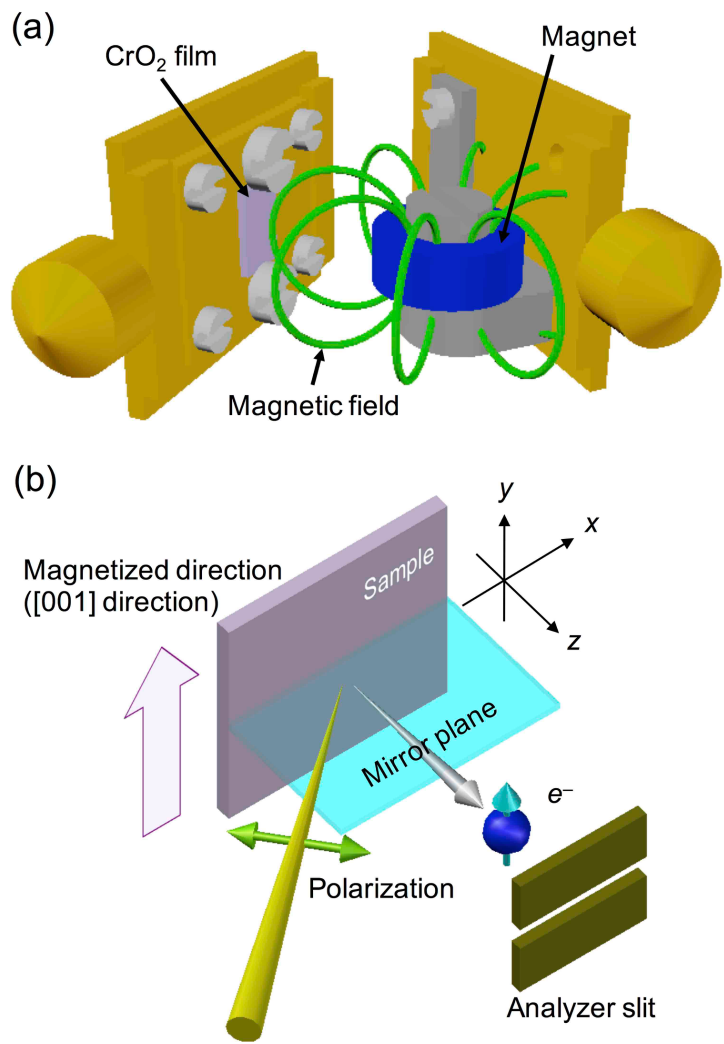

**Fig. S1 Magnetization method and experimental geometry for SRPES measurements. (a)** Schematic view illustrating how to magnetize our sample is shown. The magnet was mounted by a screw on our magnet holder. The real distance between the sample and the magnet was several millimeters, where magnitude of the magnetic field at the sample position was approximately 600 Oe. **(b)** Schematic view of the experimental geometry is represented. Both the magnetizing direction and measured spin polarization direction are parallel to the $y$ axis. The angle between the laser light and the analyzer was fixed to 50°. The acceptance angle of the spin detector was 0.7° along $x$ direction.

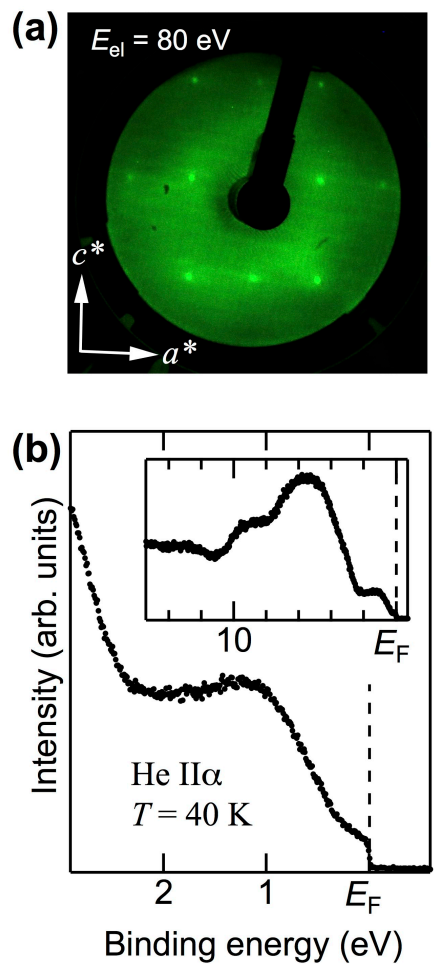

**Fig. S2 Surface quality of our $CrO_2$ sample.** (a) LEED pattern at an incident electron energy of 80 eV measured after SRPES measurements. (b) Valence band spin-integrated spectrum measured by the He IIα resonance line ($hv = 40.8$ eV) at 40 K just before SRPES measurements. Inset shows the whole valence band spectrum.

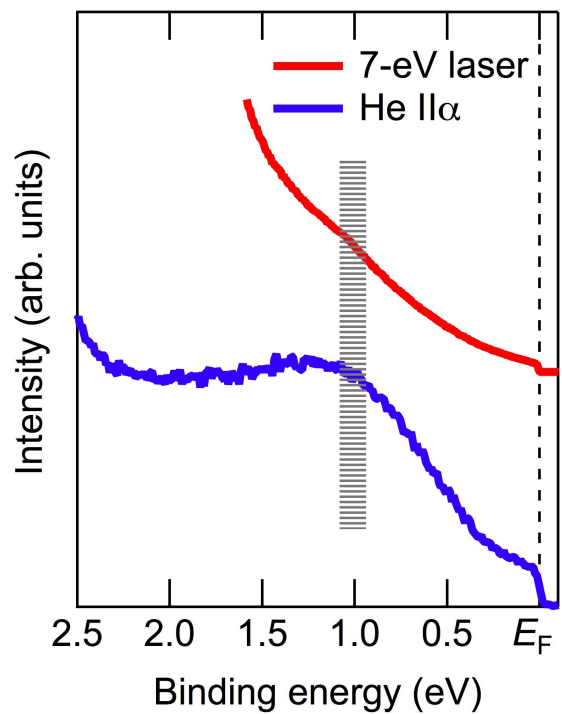

**Fig. S3 Photon-energy dependence of angle-integrated PES spectra.** Gray horizontal-striped area marks a structure observed by the all photon energies. There is a spectral structure located at 1 eV, characteristics of $d_{xy}$ band of $CrO_2$, in all spectra.

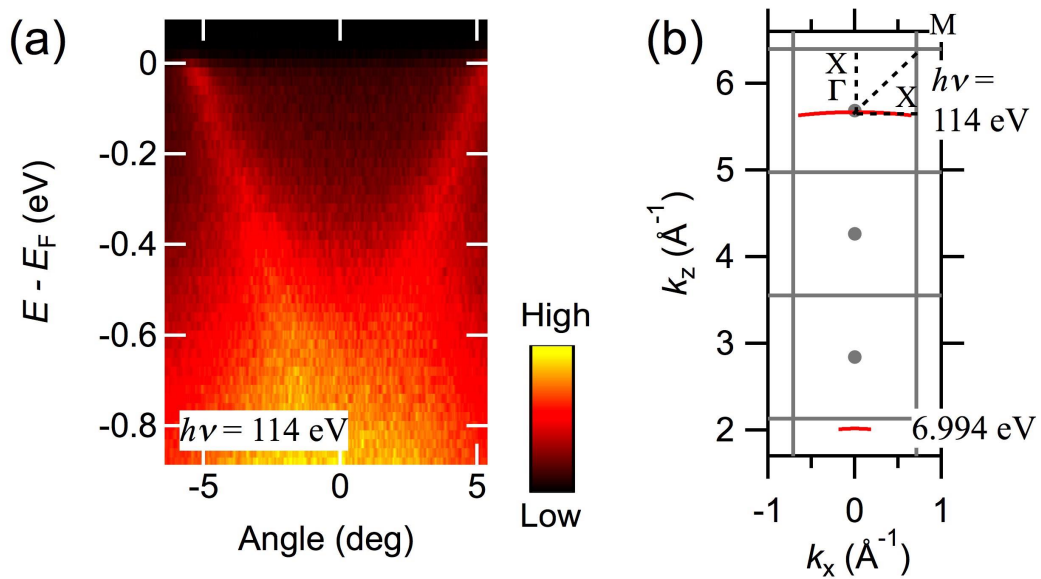

**Fig. S4 Photon-energy dependence of ARPES intensity map.** (a) ARPES intensity map measured by $h\nu = 114$ eV. The temperature was set below 20 K. (d) Two-dimensional Brillouin zone showing estimated measured $k$ positions for the states near $E_F$ (red curves).

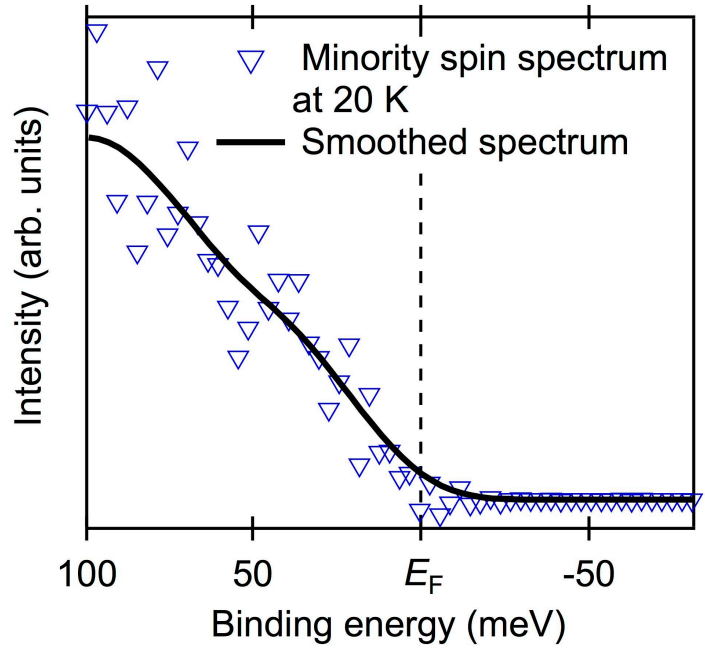

**Fig. S5 Minority spin spectrum at 20 K and its smoothed spectrum.** We used the smoothed spectrum as a background in the majority and minority spin spectra at the measured temperatures.

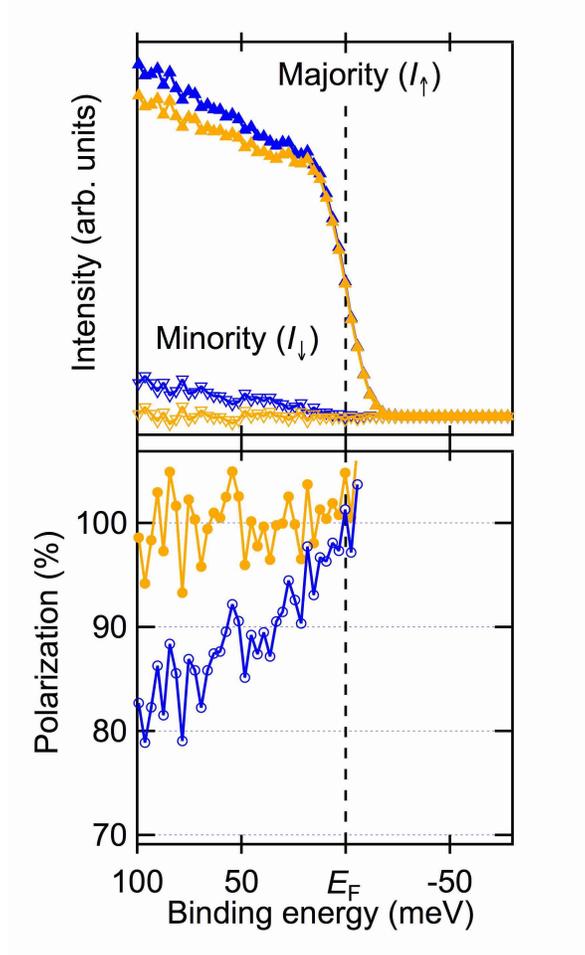

**Fig. S6 SRPES spectra and spin polarization before and after subtracting the background.**
Majority (triangle-up) and minority (triangle-down) spin spectra before (blue) and after (yellow) subtracting the background (Top), and corresponding spin polarization (Bottom).

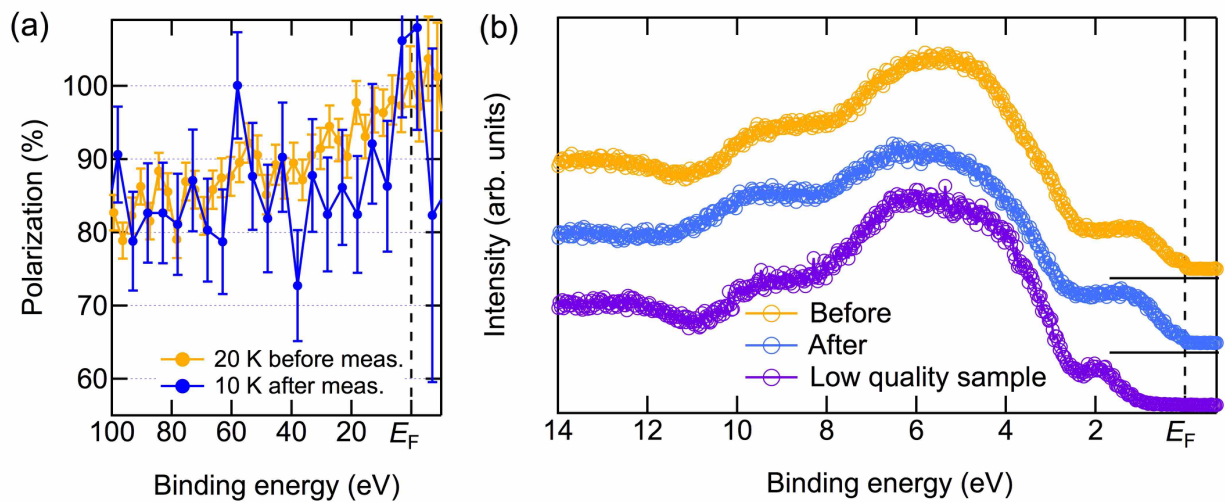

**Fig. S7 Reproducibility of spin polarization and aging check of our sample.** **(a)** Spin polarizations before (yellow) and after (blue) SRPES measurements at low temperatures. **(b)** Valence band photoemission spectra before (yellow) and after (light blue) SRPES measurements. Purple line with empty circles shows the spectrum of a low-quality sample of $CrO_2$ film whose surface seems to be almost covered by $Cr_2O_3$-like contaminants. (a) and (b) do not show significant aging near $E_F$ while the SRPES measurement.

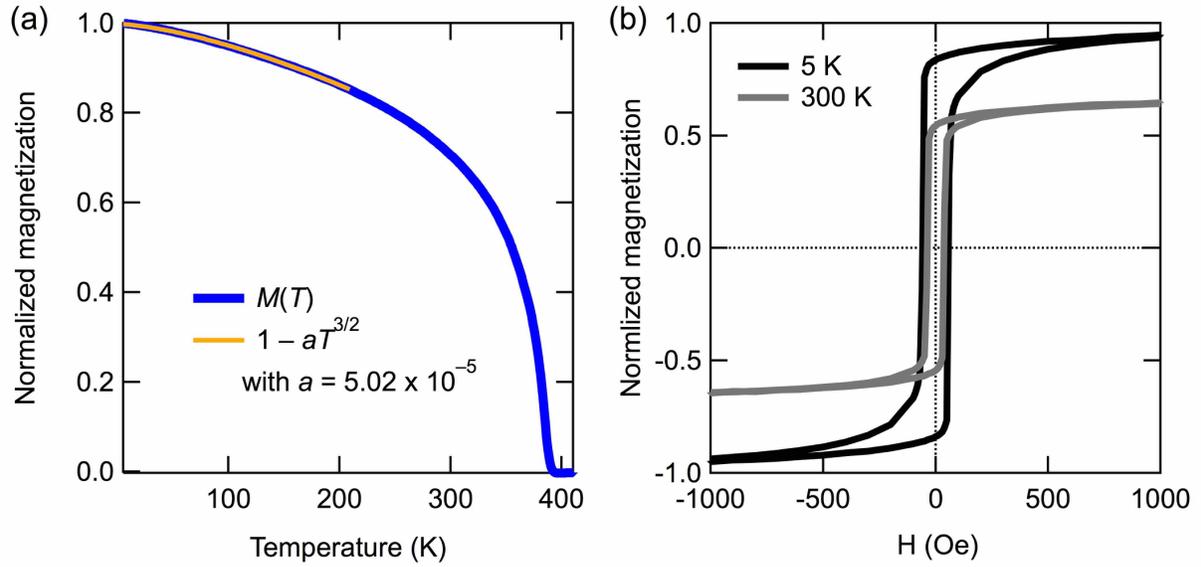

**Fig. S8 Magnetic properties of our CrO$_2$ sample. (a)** Normalized remnant magnetization $M(T)$ along *c*-axis direction for the CrO$_2$ sample which was magnetized by a magnetic field of 1 T at 300 K (blue) and the fitting curve represented by $1 - aT^{3/2}$ (Bloch's law) with $a = 5.02 \times 10^{-5}$ (yellow). **(b)** Magnetic hysteresis curves measured at 5 K (black) and 300 K (gray) which is normalized by the value at 5 K under magnetic field of 1 T.